\documentclass[conference,compsoc]{IEEEtran}

\usepackage{graphicx}

\ifCLASSOPTIONcompsoc
  \usepackage[nocompress]{cite}
\else
  \usepackage{cite}
\fi

\usepackage{balance}  
\usepackage{multirow}
\usepackage{hhline}

\begin{document}

\title{A Multimodal Emotion Sensing Platform for Building Emotion-Aware Applications}

\author{\IEEEauthorblockN{Daniel McDuff, Kael Rowan, Piali Choudhury, Jessica Wolk, ThuVan Pham and Mary Czerwinski}
\IEEEauthorblockA{Microsoft Research\\
Redmond\\
USA}}


%


\maketitle

\begin{abstract}
Humans use a host of signals to infer the emotional state of others. In general, computer systems that leverage signals from multiple modalities will be more robust and accurate in the same task. We present a multimodal affect and context sensing platform. The system is composed of video, audio and application analysis pipelines that leverage ubiquitous sensors (camera and microphone) to log and broadcast emotion data in real-time.  The platform is designed to enable easy prototyping of novel computer interfaces that sense, respond and adapt to human emotion. This paper describes the different audio, visual and application processing components and explains how the data is stored and/or broadcast for other applications to consume. We hope that this platform helps advance the state-of-the-art in affective computing by enabling development of novel human-computer interfaces.
\end{abstract}


%
\IEEEpeerreviewmaketitle

\section{Introduction}
Emotions and expressions play a significant role in our daily lives. They influence memory~\cite{miranda2005mood}, decision-making~\cite{loewenstein2003role} and social communication. Non-verbal and verbal signals (e.g., speech, facial expressions, physiological responses, gestures, language) capture rich information about these affective states. 
Human-computer interactions are rich with emotional information that is not currently considered by conventional computer systems. Thanks to the development of miniaturized electronics, ubiquitous devices are now equipped with hardware that can be leveraged to sense human affective states~\cite{calvo2015oxford}. Such sensing can help us understand human-human interactions and human-computer interactions and design computers that more fluidly and naturally interface with people. Emotion-aware devices and services have a huge potential for new assistive devices~\cite{mcduff2018designing}.  

Expressions of emotion often manifest multimodally. For example, laughter frequently occurs with smiles, and frustration with an elevated pulse rate. Thus recognition of visual, auditory and physiological signals can provide a more complete picture of the emotional state of a person than a single modality alone~\cite{d2015review}. Building multimodal systems that are able to detect expressions of affect is non-trivial. Such systems require expertise in computer vision, audio processing, natural language processing and psychology. Once models are built there is considerable software engineering effort required to build real-time applications that successfully synchronize these data.
Emotion-sensing software is a prerequisite for creating emotion-aware systems. To build such a platform for every new instance of an emotion-aware application has numerous drawbacks. First, it requires considerable effort and expertise. Much of this work can involve creating components that are not novel and simply ``re-inventing the wheel". Second, it makes it difficult to compare across systems and draw conclusions from meta-analyses if different emotion-sensing components are used each time. There is a non-negligible advantage in using a consistent set of algorithms for sensing.

\begin{figure}[!t]
\centering
\includegraphics[width=\linewidth]{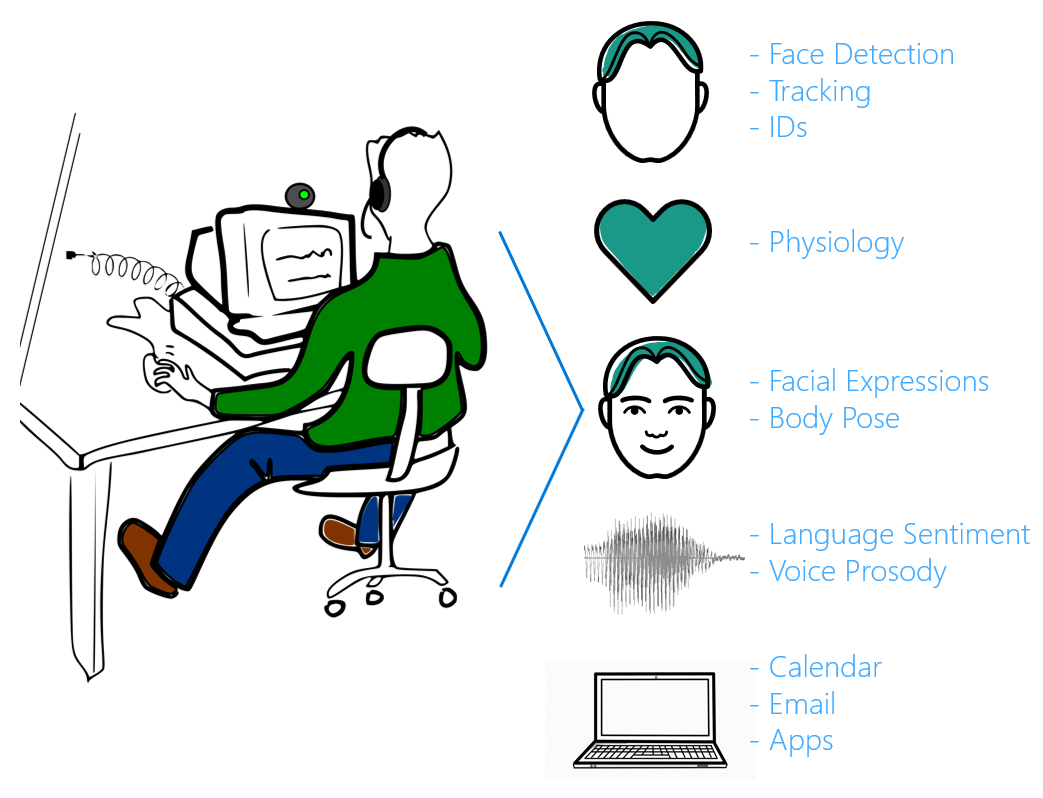}
\caption{We have developed a real-time, multimodal platform for sensing affect. The platform is designed to be flexible and extensible allowing inputs from one or multiple sensors. The platform leverages ubiquitous sensors (webcam and microphone) and has both visual and audio processing pipelines. The sensing components include face detection, recognition and tracking, expression analysis, voice activity detection, speech recognition and sentiment analysis.}
\label{fig:summary}
\end{figure}

Multimodal affect sensing platforms, such as Multisense~\cite{stratou2017multisense}, have proven effective at enabling the development of more complex human-computer interaction paradigms (e.g., SimSensei~\cite{devault2014simsensei}).
These platforms make it easier to develop applications that benefit from affect sensing capabilities.
We have developed a real-time, multimodal platform for sensing affect. The platform is designed to be flexible and extensible allowing inputs from one or multiple sensors (see Figure~\ref{fig:summary} for an illustration of the sensing modalities). The platform leverages ubiquitous sensors (webcam and microphone) and has both visual and audio processing pipelines.  Components enable face detection, recognition and tracking and expression analysis, pose estimation, voice activity detection, speech recognition and sentiment analysis. To provide context we also include application logging components that monitor application usage, email and calendar activity. Combined these data provide a rich picture of the behavior, expressions and activities of the subject(s).

There are trade-offs between designing a sensing platform that runs on-device or on a cloud server.  Platforms that run on-device have the advantage of processing raw images and audio sensor data locally and minimizing the transmission of Personally Identifiable Information (PII). They remove the need for network connectivity. Cloud platforms have the advantage of less constraints on computational resources. The platform can be tailored to the hardware available and allow for greater scalability. However, cloud systems typically suffer from greater latency and the need to stream raw data over a network. Our platform can be run on Windows 10 devices locally or on a cloud virtual machine (VM).  This enables the ability to create native and browser-based applications.

We hope our platform facilitates the measurement and understanding of affective signals and enables research into how these can be integrated into a wide range of services.  In the remainder of the paper, we describe the architecture, sensing components, data management and data storage. First, we start with a brief review of some related work.

\section{Related Work}
There is an extensive literature on automated affect recognition. We will not cover the prior work completely here as surveys of the existing work provide a much more complete review of the field than would be possible in this section~\cite{d2015review,soleymani2017survey}. However, we highlight a few highly relevant and seminal papers on multimodal affect recognition platforms and applications.

Multimodal affect sensing has been applied in numerous contexts including teaching and learning environments~\cite{kapoor2005multimodal,d2010multimodal}, healthcare~\cite{devault2014simsensei}, the arts~\cite{camurri2000eyesweb}, and human-robot interaction~\cite{alonso2013multimodal}. The first work on affect recognition started almost three decades ago where physiological sensors, cameras and microphones were used to detect a host of affective responses. Early multimodal systems often comprised of bulky equipment and wired sensors~\cite{kapoor2005multimodal}. The miniturization of electronics and improvements in wireless communications now mean that sensing can be performed more easily using off-the-shelf devices that are small and ubiquitous (such as webcams, microphones, accelerometers). 

Multisense~\cite{stratou2017multisense} is a platform for multimodal affect sensing that incorporates both visual and audio components. Specifically, components included 3D head position-orientation and facial tracking, facial expression and gaze analysis, and audio analysis. It leverages existing public tools for some of these components. For example, audio analysis is performed using the OpenSmile~\cite{eyben2010opensmile} package. SimSensei~\cite{devault2014simsensei} is a virtual human interviewer designed to create engaging face-to-face interactions that are driven in part via the Multisense sensing algorithms. Multisense broadcasts signals to the Kiosk using the Physical Markup Language (PML) standard. 

The Platform for Situated Intelligence (PSI)~\cite{bohus2017rapid,psiblogpost} is a new open source platform for building multimodal interactive systems. One of the aims of this project was to make it easier for developers to create multi-sensor and multimodal architectures and handle low-level elements such as data synchronization. While not specifically designed for affective computing applications it contains many of the elements needed in this domain.  

\section{Architecture}
\begin{figure}[!t]
\centering
\includegraphics[width=3.5in]{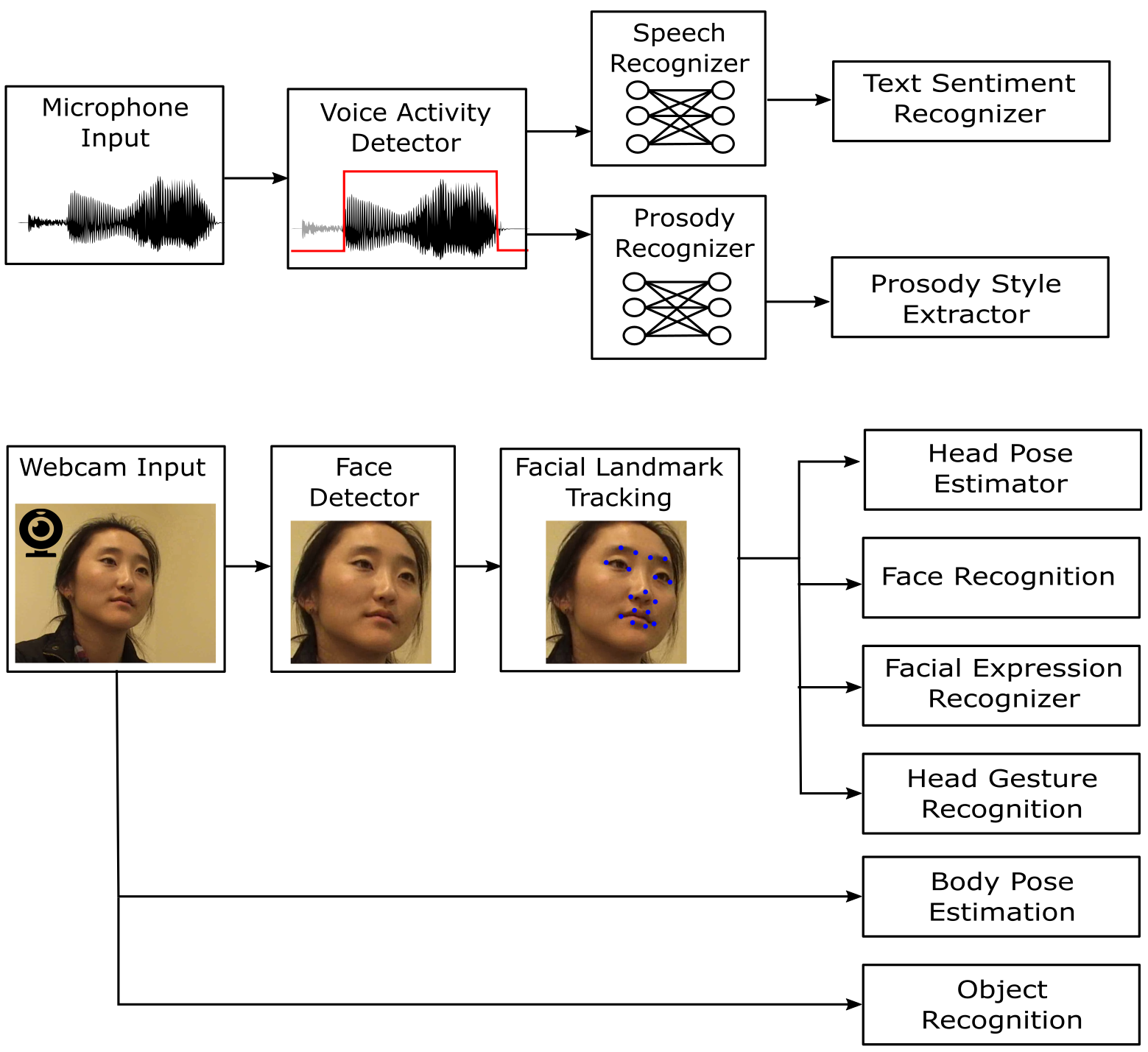}
\caption{Block diagrams of the audio and visual processing pipelines in our platform.  The components can easily be removed and/or new components added, this specific sensing system is just one example of one configuration of components.}
\label{fig:architecture}
\end{figure}

Our platform is composed of a number of processing \textit{components} that are connected together to form \textit{pipelines}.  The system is built using PSI~\cite{bohus2017rapid,psiblogpost}, from which we inherit this terminology.  Figure~\ref{fig:architecture} shows an overview of the current sensing system.  The nature of the components is that they can easily be removed and/or new components added, thus this specific sensing system is just one example of one configuration of components. 

\begin{figure*}[!t]
\centering
\includegraphics[width=\linewidth]{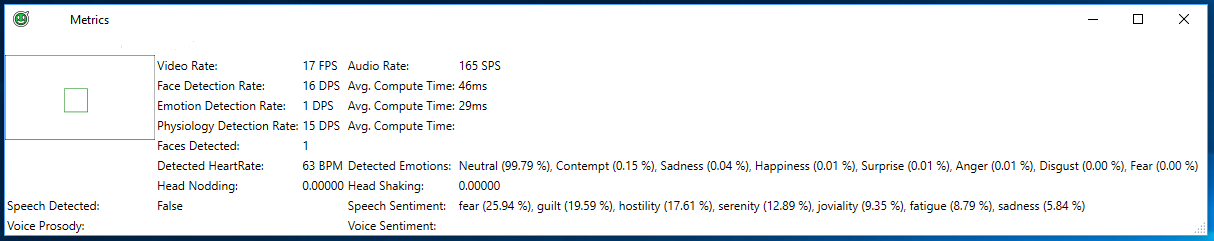}
\caption{Debugging metrics window that shows the output parameters from the different sensing components in real-time. The green box highlights to location of a face within the video frame. The video frame is not shown to highlight that no raw video (or audio) data is stored.}
\label{fig:metrics_window}
\end{figure*}

Table~\ref{table_example} shows a summary of the audio and visual components and the outputs that are typically logged by the software. To help those who develop applications using the emotion-sensing platform we created a simple debugging window, shown in Figure~\ref{fig:metrics_window}, that displays the values of these outputs in real-time. 

\section{Vision Components}

\subsection{Face Detection and Tracking}
The video signals from the webcam were sampled at 15 frames-per-second (FPS). We used the Microsoft Face API\footnote{https://azure.microsoft.com/en-us/services/cognitive-services/face/} to detect the faces in each of the video frames and apply a landmark detector to identify the eyes, nose, and mouth. We log the number of faces in the frame of the camera, the location of this faces (both bounding box and 15 landmark points). If multiple faces were present in a single frame the faces are given ids based on their location. Using a heuristic rule-based method these ids were propagated to the next frame.  Therefore, if the people did not move dramatically or leave the frame of the camera they were given the same id in the subsequent frame. More sophisticated tracking can be enabled by employing a face recognition step, as described below.
 
\subsection{Face Recognition}
We leverage a deep neural network to determine an embedding for faces detected. The embeddings are designed to be unique for each individual and robust to changes in appearance (e.g., hair style and clothing) or the environment. This face embeddings can be used to supersede the heuristic rule-based tracking if the face recognition component is employed.

\subsection{Facial Expression Recognition}
The faces are cropped from the video frames using the bounding box information provided by the face detector and the resulting image-patches sent to an facial expression detection algorithm.  The facial expression detector returns eight probabilities, one for each of the following basic emotional expressions: anger, disgust, fear, happiness, sadness, surprise and neutral.  This is a frequently employed categorization of facial expressions; however, it is not without critics and displays of emotion are not uni-modal or necessarily universal~\cite{jack2012facial}. Nevertheless researchers have successfully built algorithms to identify useful signals in the noisy world of human expression and that these signals are consistent with socio-cultural norms~\cite{mcduff2017large}.  We used the publicly available EmotionAPI\footnote{Microsoft, Inc.} emotion detector, allowing other researchers to replicate our method. The emotion detection algorithm is a convolutional neural network (CNN) based on VGG-13, more details can be found in~\cite{barsoum2016training}. We log the emotion probabilities for each face detected for each frame that is processed. The emotion detector can be run at less than 15 FPS in order to reduce the computational load of the software.  We validated the facial affect classification on two independent public benchmark datasets, the CK+ dataset~\cite{lucey2010extended} and the FER dataset testing set~\cite{goodfellow2013challenges} (using the FER+ labels~\cite{barsoum2016training}). These datasets are comprised of 326 and 3,573 labeled images of the same basic emotion categories as our classifier, respectively. To characterize performance we report the true positive rate, false positive rate and accuracy for the task of categorizing the facial expression images.  For CK+ the accuracy is 81.0\%, the false positive rate is 3.04\% and the true positive rate is 72.5\%. For FER the accuracy is 82.0\%, the false positive rate is 2.94\% and the true positive rate is 76.1\%.
While facial expressions alone will not give a complete picture of an individual emotional state, these results reflect that the model is able to detect facial expressions of emotion with reasonable reliability and gives relatively consistent performance across two different datasets. 

\subsection{Body Pose Estimation}
We use a video-based body pose estimation method based on the approach described in~\cite{xiao2018simple}. This is a 2D pose estimation algorithm based on RGB video input. The algorithm is trained on the COCO dataset~\cite{lin2014microsoft}. The component returns the location of keypoints on the body which are mapped to major joints (e.g., knees, shoulders, elbows, wrists, etc.).  The component uses a ResNet-based~\cite{he2016deep} approach implemented in PyTorch.

\subsection{Imaging PPG}
We use an approach for imaging photoplethysmography (iPPG) to extract heart rate estimates from the faces within the webcam video feed. The iPPG algorithm is implemented in a manner similar to the method presented in~\cite{poh2011advancements,mcduff2014improvements}. The face detection bounding box is used to define the region of interest (ROI) of the face in each frame. The RGB color signals in this ROI are spatially averaged for each frame and a window of length 300 frames ($\sim$ 20 seconds) is constructed from the spatially averaged signals. These signals are then detrended and used as input to an Independent Component Analysis (ICA). The resulting source signals are analyzed in the frequency domain (using Fast Fourier Transforms) to determine the estimated HR. See~\cite{mcduff2014improvements} for more details. The method relies on observations from the face being complete for 300 frames in order to make an estimate; therefore, there is a short delay between when a face is detected and when as HR estimate is reported. We log the HR estimate for each 300 frame window. Previous, validations of this approach have found accuracy to be within one to two beat-per-minute (BPM) mean absolute error when subjects are well lit and stationary~\cite{poh2011advancements,blackford2015effects}.
Respiration rate can be estimated using a similar approach, by analyzing the RGB values across time but employing different filtering parameters and fitting an auto-regressive model and performing pole selection~\cite{tarassenko2014non}. For a survey of techniques see~\cite{mcduff2015survey}.

\begin{table}[!t]
\renewcommand{\arraystretch}{1.3}
\caption{Summary of the detection and recognition components in our platform. The specific parameters that are logged/broadcast are indicated.}
\label{table_example}
\centering
\begin{tabular}{rll}
\hline 
 & Component & Outputs \\
\hline \hline
\multirow{4}{*}{\rotatebox[origin=c]{90}{Microphone}} & Voice Activity Detection (VAD) & Boolean (VAD true/false) \\
 & Speech Recognition & String: Text \\
 & Language Sentiment Analysis & Floats: Probs. of 8 emotions \\
 & Voice Prosody Analysis & Floats: Pitch, Energy \\ \hline
\multirow{4}{*}{\rotatebox[origin=c]{90}{Camera}} & Face Detection & Ints: Bounding box locations \\
 & Face Recognition & Ints: Face embeddings \\
 & Face Expression Recognition & Floats: Probs. of 8 emotions \\
 & Pose Estimation & Floats: Joint locations \\
 & Physiology & Floats: HR and Resp. rate \\ \hline
 \multirow{4}{*}{\rotatebox[origin=c]{90}{Applications}} & Appications & Strings: Window titles \\
 & Calendar & Strings: Calendar event details \\
 & Email & Floats: Email sentiment scores \\
 & Keyboard & Booleans: Keyboard typing \\
 & Mouse & Booleans: Mouse movement \\
\hline
\end{tabular}
\end{table}

\section{Audio Components}
\subsection{Voice Activity Detection}
The audio received from the microphone is sampled at 16kHz and passed through a Voice activity detector. We use the Microsoft Windows Voice Activity Detector (VAD)~\cite{tashev2015offline}.  Speech activity is logged as a Boolean variable. In addition, to allowing speech to be segmented for the purposes of further analysis, this also provides useful contextual information about whether a subject is talking. Combined with facial detection this allow for inference of in-person social interactions.

\subsection{Speech Recognition}
Segments of audio for which voice activity is detected are sent to a cloud-based speech-to-text (STT)~\cite{MSspeechapi}. This returns the single most probably string based on the STT algorithm. This is the only part of the system that requires a cloud service, due to the superiority of the cloud-based speech recognition engines compared to the models that were available locally.

\subsection{Language Sentiment Analysis}
The strings returned from the STT engine are passed through a language sentiment analysis component.  This classifier returns probabilities for eight sentiment categories: joviality, fear, sadness, surprise, hostility, serenity, fatigue and guilt.  The classifier was trained on Twitter social media posts collected via the Twitter fire hose.  Details of the data, training and validation can be found in~\cite{de2012happy}. The sentiment classification uses somewhat similar categories to the facial expression classification, but with some differences to make the categories more appropriate for spoken language. As with facial expression categorization these classes are unlikely to be universal and are certainly not exhaustive; however, they provide useful signals.

\subsection{Voice Prosody Analysis}
We employ a deep neutral network (DNN) model to estimate the emotional valence (negative, neutral, positive) of the voice tone. First, the fundamental frequency and MEL frequency features are extracted from the voiced segments of audio.  Then a DNN model is used to estimate the probabilities of the the three emotional valence categories. The scores for the three classes (negative, neutral, positive) sum to one.

\section{Application Logging Components}
Often when tracking non-verbal expression and emotion data it is difficult to analyze or interpret the observations effectively without context on what the subject is doing. When running locally on a client our tool has components for tracking application usage, calendar events and email information.

\subsection{Application Logging}
The application logging captures the applications running on the machine. For each application we log the name, position on the screen and events (starting and closing the application, minimizing and maximizing the window, bringing the window to the foreground). 

\subsection{Email Logging}
The body of each email that the participants sent is classified using a textual sentiment classifier to extract emotional valence (positive and negative). The classification algorithm used is a logistic regression trained on a set of text corpora from multiple domains labeled for sentiment~\cite{pang2004sentimental,blitzer2007biographies,ganapathibhotla2008identifying} and an internal dataset of labeled comments and reviews. The textual features are unigrams (single words) and bigrams (pairs of words in sequence). During training hyperparameter tuning was performed via a grid search of parameters and five-fold cross-validation. Feature selection was performed by using mutual information criteria to select the top 5,000 unigrams or bigrams, and domain specific features were manually pruned to leave 1,200 features.

\subsection{Calendar Logging}
One of the richest sources of contextual data stored digitally is a person's digital calendar. Our software logged details of calendar events from Microsoft Outlook, including the number of attendees, start time, duration and whether the meeting was in-person or via Skype.

\section{Local versus Cloud}
Our platform allows components to be run locally (on device) or in the cloud. If run in the cloud the video and audio are streamed via a communication protocol (i.e., WebRTC). Running the platform locally allows for minimal transmission of PII and for most components to be used even when there is no Internet connection available.  Running the platform in the cloud allows greater scalability and exploitation of greater computational resources than are available on the device. In the simplest form this only requires the participant to have a device that can run a web browser and support a webcam and microphone (e.g., smartphone, tablet, Raspberry Pi etc.).


\section{Logging and Broadcasting}
The system was designed with two data interfaces. First, the data is logged to a secure cloud server. The output from each component is averaged (or appended in the case of strings) over an aggregation window (default length one second). These values are stored in a single row of the cloud database.  Second, the data is broadcast over a secure service bus that allows other applications, with permission, to ``listen" for the data. This makes it easy to build applications on top of the sensing framework without having to interface with the code itself. The broadcast data can be read easily using Python and Javascript. To date, prototype applications have been developed using each of these interfaces.

\section{Consent}

Collecting rich information about behavior, expressions and applications raises important questions about privacy. When installed on personal computers the sensing platform is always run with explicit consent from the users.  On installation they are presented with a consent form that details the information that will be logged. No raw audio, video or textual data is stored, including no titles or bodies of email messages, titles of calendar events or names of attendees of calendar events.  The participants are always informed about how to start and stop the software and are told they are free to do so.
The names, email addresses or other clearly identifiable information about the participants are also not linked to the data.

\section{Conclusion}

Non-verbal and verbal signals capture rich information about affect. Ubiquitous devices are now equipped with sensors that can be leveraged to quantify affective and emotional states.  Multimodal sensing technology is non-trivial to build, such systems require expertise in computer vision, audio processing, natural language processing and psychology. 

To help expedite the development of emotion-aware systems we developed a multimodal sensing platform that is designed to run on device or in the cloud. It includes components for face detection, recognition and tracking and expression analysis, pose estimation, voice activity detection, speech recognition and sentiment analysis.  The on device option allows data to be processed locally minimizing the need to transmit PII. The cloud option reduces constraints on the locally available computational resources. 

We hope that this platform is transparent and extensible and that we can create interfaces and experiences that build on top of it without ``reinventing" affect sensing components every time.

\ifCLASSOPTIONcompsoc
  \section*{Acknowledgments}
\else
  \section*{Acknowledgment}
\fi

The authors would like to thank Michael Gamon, Mark Encarnacion, Ivan Tashev, Cha Zhang, Emad Barsoum, Dan Bohus and Nick Saw for the contribution of models and PSI components that are used in this platform.

\balance{}



\bibliographystyle{IEEEtran}
\bibliography{references}
%



\end{document}